\documentclass[10pt,a4paper]{article}
\usepackage{amsmath}
\usepackage{amsfonts}
\usepackage{amssymb}
\usepackage{graphicx}
\usepackage{textcomp}
\usepackage{url}
\usepackage{lipsum}
\usepackage{setspace}
\usepackage{comment}
\usepackage{color}
\usepackage[numbers,sort&compress]{natbib}
\usepackage[usenames,dvipsnames,svgnames,table]{xcolor}
\usepackage[lofdepth,lotdepth,caption=false]{subfig}
\usepackage[pdfstartview=FitH]{hyperref}
\usepackage{hyperref}
\newsavebox{\bigleftbox}
\hypersetup{
 colorlinks,
 citecolor=Blue,
 linkcolor=Blue,
 urlcolor=Blue
}

\makeatletter
 \def\footnoterule{\kern-3\p@
   \noindent\hrulefill \kern 2.8\p@} 
\makeatother
\usepackage[left=3.00cm, right=3.00cm, top=2.00cm, bottom=2.00cm]{geometry}

\usepackage{fancyhdr}

\title{\textbf{Unveiling a Novel Silicene-Like Material: A DFT Study on Pentahexoctite-Silicon and Its Optoelectronic Characteristics}}
\author{
    K. A. L. Lima$^{1,2,\P}$, 
	L. A. Ribeiro Junior$^{1,2,\S}$
	}
\date{}

\begin{document}
    \maketitle
	\vspace{-0.6cm}
	\begin{center}\small
	\textit{Institute of Physics, University of Bras\'ilia, 70910-900, Bras\'ilia, Brazil}\\
            \textit{Computational Materials Laboratory, LCCMat, Institute of Physics, University of Bras\'ilia, 70910-900, Bras\'ilia, Brazil}\\
		\phantom{.}\\ \hfill
        $^{\P}$\url{kleutonantunes@gmail.com}\hfill
		$^{\S}$\url{ribeirojr@unb.br}\hfill
		\phantom{.}
	\end{center}
	

\onehalfspace

\noindent\textbf{Abstract: Silicon-based two-dimensional (2D) materials, including well-known silicene, have garnered considerable attention due to their potential in advanced electronic and optoelectronic applications. Here, we introduce a novel 2D silicon variant, pentahexoctite silicon (PH-Si), inspired by the unique structural attributes of pentahexoctite carbon. By using state-of-the-art density functional theory (DFT) calculations and ab initio molecular dynamics (AIMD) simulations, we investigated the fundamental optoelectronic traits of PH-Si. Our findings unveil that PH-Si boasts promising electronic properties characterized by anisotropic conductance and metallic behavior along specific directions. We also examine its structural and dynamic stability through phonon calculations and AIMD simulations. PH-Si structural stability was also confirmed by its low formation energy of -3.62 eV. The material exhibits substantial optical absorption in visible (Vis) and ultraviolet (UV) spectral regions, positioning it as a potential Vis-UV detector and absorber. Additionally, we assess its crucial mechanical attributes, encompassing elastic stiffness constants, Young's modulus (ranging from 5 to 40 GPa), and a Poisson ratio of 0.8, collectively offering valuable insights into its mechanical performance.}

\section{Introduction}

Silicon, the second most abundant element on Earth's crust, has proven to be an indispensable component in the modern electronics industry \cite{hwang2012physically,zwanenburg2013silicon}, serving as the fundamental material for semiconductor devices \cite{del2011nanometre}. The quest for novel materials to expand the horizons of silicon-based technologies has led to the exploration of new two-dimensional (2D) materials \cite{teo2007silicon}. Silicene \cite{balendhran2015elemental,kara2012review,zhao2016rise,PhysRevLett.102.236804}, an atom-thin silicon allotrope with a honeycomb lattice similar to graphene, has garnered substantial attention due to its intriguing electronic properties and potential applications in nanoelectronics \cite{pan2015silicene,zhao2016rise} and photovoltaics \cite{lv2018direct}. However, the practical utility of silicene could be improved by the challenges of synthesis and stability \cite{jose2014structures}, calling for alternative silicon/silicene-based 2D materials \cite{molle2018silicene,man2023two,masson2023epitaxial} with distinct lattice arrangements and enhanced properties.

Recent advances in material synthesis have produced intriguing 2D silicene allotropes \cite{sheverdyaeva2017electronic,stpniak2019planar}. The primary objective is to fabricate novel allotropes capable of addressing the inherent zero bandgap limitation of conventional silicene. This limitation currently restricts its utility in specific optoelectronic devices that demand materials with semiconducting bandgaps. In this way, a combination of experimental observations using angle-resolved photoemission and density functional theory (DFT) calculations has unveiled the existence of various silicene-based structures, each with its unique structural characteristics. Despite their structural variations, these silicene allotropes exhibited similar electronic band structures \cite{sheverdyaeva2017electronic}. 

In addition, scanning tunneling microscopy measurements have obtained atomically precise images of a honeycomb silicene planar lattice \cite{stpniak2019planar}, representing a new allotrope concerning conventional silicone. DFT calculations support these experimental findings and predict a pure sp$^2$ atomic configuration for silicon atoms in planar silicene \cite{stpniak2019planar}. A remarkable feature of planar silicene is its nearly flat geometry, similar to graphene's atomic structure, distinguishing it from the more common buckled silicene allotropes. 

Theoretical investigations have also expanded our understanding of potentially stable silicene allotropes, often called silicene derivatives \cite{sang2021semiconducting,qiao2017tetra,lin2017stability,tang2017ab,qian2021pc,wu2017prediction,gorkan2022functional}. Through DFT calculations, these studies have revealed the remarkable stability and semiconducting characteristics of novel silicene variants. These allotropes generally exhibit an electronic bandgap on par with bulk silicon and boast electron mobility levels comparable to those of honeycomb silicene. In particular, the lattice structures of most of these novel silicene derivatives are derived from other 2D carbon materials. This approach opens up exciting possibilities for exploring the properties of silicon-based 2D materials that inherit their structural features from established carbon counterparts.

One particularly intriguing computational carbon allotrope that has attracted considerable attention is pentahexoctite \cite{sharma2014pentahexoctite}. This 2D material exhibits a unique sp$^2$ hybridized structure composed of continuous 5-6-8 carbon atom rings. Electronically, pentahexoctite displays metallic properties, featuring flat and dispersive bands at the Fermi level that exhibit direction-dependent behavior, resulting in highly anisotropic transport characteristics.

The band structure of pentahexoctite represents a distinctive combination of flat and highly dispersive bands, leading to its overall metallic nature. This variation in band behavior along different directions contributes to the sheet's electronic transport anisotropy \cite{sharma2014pentahexoctite}. Moreover, pentahexoctite serves as a precursor for stable 1D nanotubes, and its electronic and mechanical properties depend on chirality, adding to its unique appeal. Given these remarkable characteristics, exploring its silicon counterpart's electronic and structural properties, referred to as PH-Si, could significantly advance our understanding of the potential impact of materials with similar topological features in various optoelectronic applications. 

In this study, we contribute to the ongoing exploration of silicon-based 2D materials by proposing PH-Si, a silicon allotrope pentahexoctite \cite{sharma2014pentahexoctite}. Our study broadens the class of 2D silicon materials, offering an alternative with potential advantages over silicene and other derivatives. By employing state-of-the-art DFT calculations and ab initio molecular dynamics (AIMD) simulations, we investigate the mechanical, electronic, and optical properties of PH-Si, shedding light on its underlying physical properties.

\section{Methodology}

\begin{figure}[!htb]
    \centering
    \includegraphics[width=0.8\linewidth]{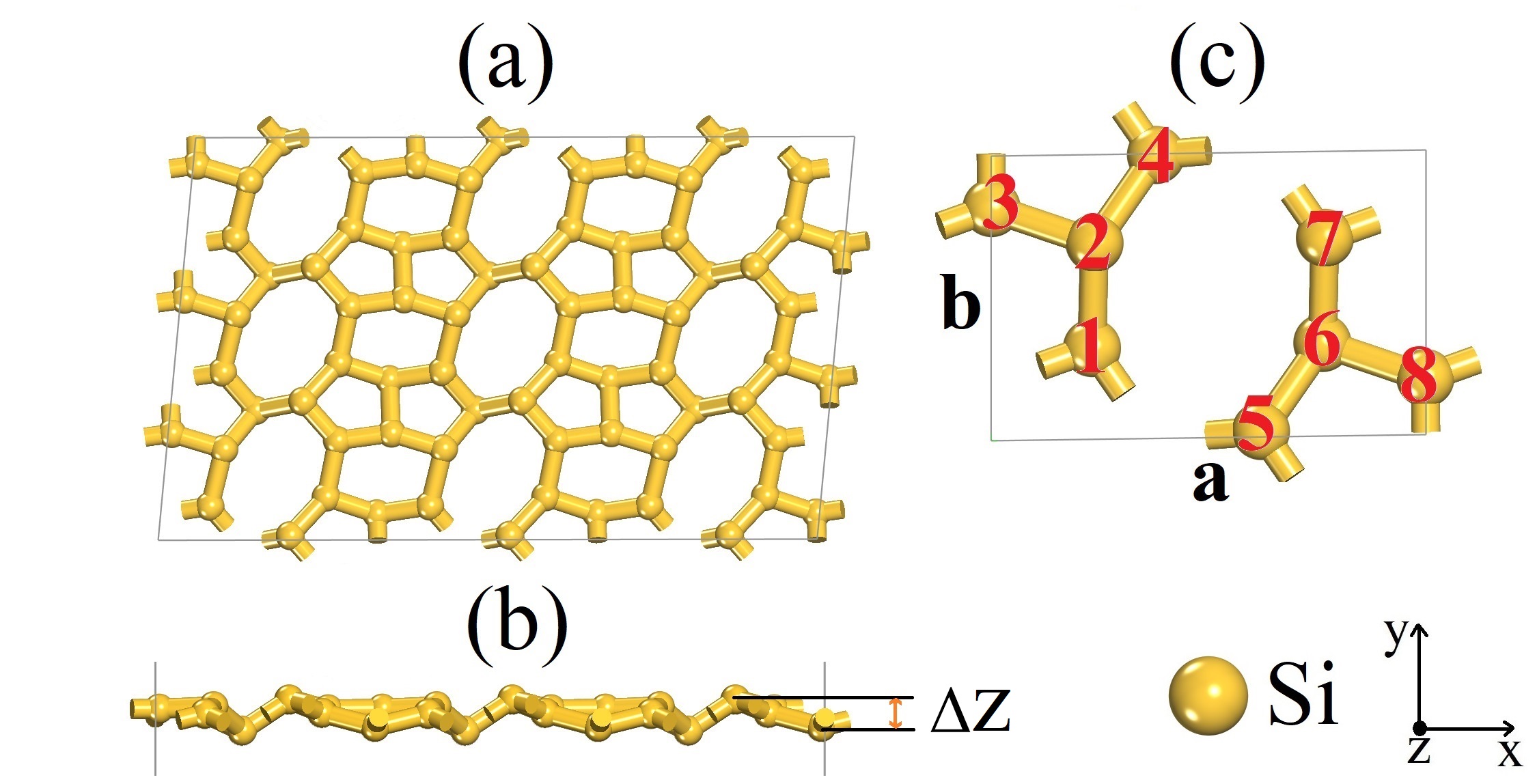}
    \caption{Schematic representation of the PH-Si sheet in the (a) top and (b) side views. The black rectangle defined by lattice vectors $a$ and $b$ highlights the unit cell.}
    \label{fig:sys}
\end{figure}

We conducted systematic DFT-based investigations into the mechanical, electronic, and optical characteristics of PH-Si. Its lattice arrangement is depicted in Figure \ref{fig:sys}. These investigations were carried out using the CASTEP code \cite{clark2005first}. In our calculations, we adopted the Generalized Gradient Approximation (GGA) to account for exchange and correlation effects. Specifically, we used the Perdew-Burke-Ernzerhof (PBE) functional \cite{perdew1996generalized} along with the Heyd-Scuseria-Ernzerhof (HSE06) hybrid functional \cite{heyd2003hybrid} to improve the accuracy of our simulations. A norm-conserving pseudopotential tailored for silicon was used, as available in the CASTEP framework.

To achieve electronic self-consistency, we used the Broyden–Fletcher–Goldfarb–Shannon (BFGS) unrestricted nonlinear iterative algorithm, as outlined in the works of Head and Ceperley \cite{head1985broyden,PFROMMER1997233}. Our calculations were conducted with a plane-wave basis set, employing an energy cutoff of 700 eV. We enforced a stringent convergence criterion for the energy, set at 1.0 $\times 10^{-5}$ eV. We also imposed periodic boundary conditions to ensure the total relaxation of the PH-Si lattice. These conditions guaranteed that the residual forces acting on each atom remained below 1.0 $\times 10^{-3}$ eV/\r{A} and that the pressure did not exceed 0.01 GPa. 

The basis vector in the z direction was fixed during the lattice optimization, using a k-grid of $10\times10\times1$ for this purpose. We used k-grids of $15\times15\times1$ and $5\times5\times1$ for the PBE and HSE06 methods for electronic and optical computations, respectively. To analyze the Partial Density of States (PDOS) at the HSE06 level, we consider a k grid of $20\times20\times1$. A vacuum region of $20$ \r{A} was employed to avoid spurious interactions among periodic images. For deriving elastic properties, we applied the LDA/CA-PZ method \cite{PhysRevLett.45.566,PhysRevB.23.5048}. 

We adopted the linear response methodology for phonon calculations with a grid spacing of 0.05 \r{A}$^{-1}$. We set a convergence tolerance of 10$^{-5}$ eV/\r{A}$^{2}$. In evaluating the mechanical properties of the PH-Si, we employed the stress-strain method, which relies on the Voigt–Reuss–Hill approximation \cite{Zuo:gl0256,10.1063/1.1709944}.

Our study also involved the computation of the effective mass ($m^*$) for both electrons and holes, a fundamental parameter for determining the overall carrier mobility in 2D materials. This computation entailed fitting the band dispersion to the appropriate models to derive this critical parameter as follows,

\begin{equation}
 \displaystyle m^{*}=\hslash^2\left(\frac{\partial^2E(k)}{\partial k^2}\right)^{-1}.   
\end{equation}

\section{Results}

Figure \ref{fig:sys} illustrates the optimized lattice of PH-Si. The geometry optimization calculations produced consistent lattice parameters using the PBE and HSE06 methods. However, for this discussion, we emphasize the results obtained using the HSE06 method, as presented in Figure \ref{fig:sys} and Table \ref{tab:bonds}. The lattice vectors $a$ and $b$ are measured at 8.99 \r{A} and 5.88 \r{A}, respectively (as shown in Figure \ref{fig:sys}). The crystal structure of PH-Si adheres to an orthorhombic arrangement precisely aligned with the P-1 (CI-1) space group.

\begin{table}[!htb]
\caption{Bond distances for the atoms highlighted in Figure \ref{fig:sys} calculated at the HSE06 level.}
\centering
\label{tab:bonds}
\begin{tabular}{|c|c|c|c|}
\hline
Bond Type & Bond Length (\r{A}) & Bond Type & Bond Length (\r{A}) \\ \hline
S1-S2    & 2.45                          & S5-S6    & 2.25                         \\ \hline
S2-S3    & 2.25                         &  S6-S8    & 2.28                      \\ \hline
S2-S4    & 2.30                       & S6-S7    & 2.39                        \\ \hline
\end{tabular}
\end{table}

In Table \ref{tab:bonds} we present bond length values that closely resemble those observed in conventional hexagonal silicene at approximately 2.28 \r{A} \cite{PhysRevLett.102.236804} and TO-silicene \r{A} \cite{wu2017prediction}, which falls within the range of 2.25-2.30 \r{A}. These bond lengths in PH-Si suggest that this material may exhibit stability on par with silicene. The Si-Si bond lengths in PH-Si are also similar to those found in pristine pentagonal-ring-based silicon materials, such as monolayer P-silicene \cite{aierken2016first} (about 2.23-2.36 \r{A}) and single PSi-cluster \cite{sheng2018pentagonal}, with Si-Si bonds measuring 2.36 \r{A}. In addition, like other 2D silicon materials, PH-Si also exhibits buckling characteristics, with a $\Delta Z$ value (refer to Figure \ref{fig:sys}) of approximately 0.58 \r{A}. This value is similar to the 0.44 \r{A} found in silicene \cite{PhysRevLett.102.236804}. Our analysis has also unveiled a formation energy value of -3.62 eV for the PH-Si lattice, suggesting the possibility of experimental synthesis of PH-Si.

\begin{figure}[!htb]
	\centering
	\includegraphics[width=0.8\linewidth]{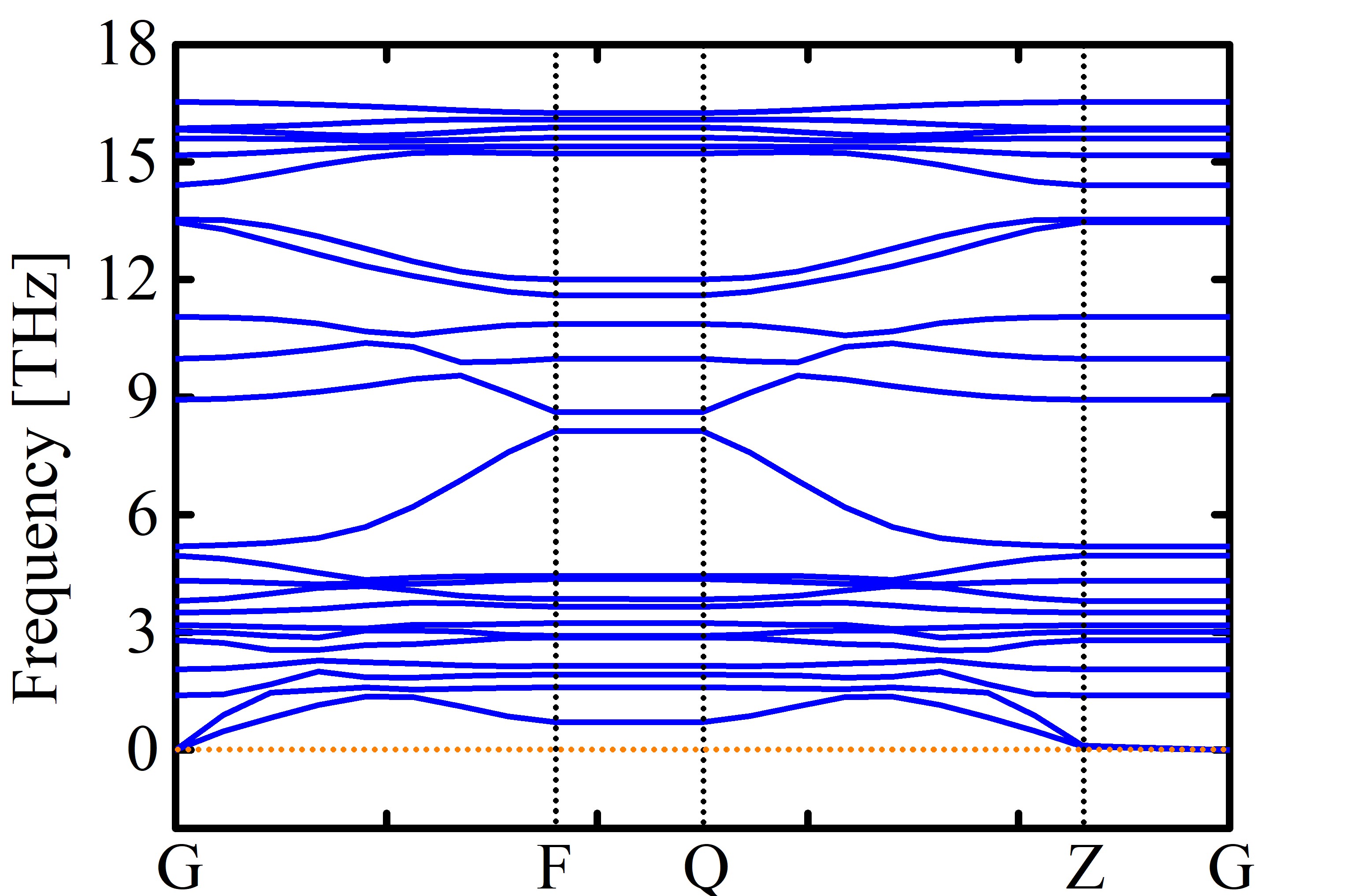}
	\caption{Phonon band structure of PH-Si calculated at PBE level.}
	\label{fig-phonons}
\end{figure}

To assess the dynamic stability of PH-Si, we calculate its phonon dispersion with the corresponding curves, as shown in Figure \ref{fig-phonons}. This figure reveals the absence of imaginary frequencies, suggesting the dynamic stability of PH-Si. The lack of a band gap between acoustic and optical modes implies a noteworthy scattering rate and shorter phonon lifetimes, contributing to the material's relatively moderate lattice thermal conductivity. 

It is a well-established principle that higher phonon frequencies indicate stronger chemical bonds. In the context of PH-Si, the highest phonon frequency is approximately 16.53 THz, slightly lower than the 17.98 THz observed in silicene \cite{PhysRevLett.102.236804}. Remarkably, despite their distinct structural topologies, PH-Si and silicene tend to exhibit Si-Si bonds with comparable bond energies. In this case, the Si-Si bonds in PH-Si are marginally weaker than those in silicene.

In Figure \ref{fig-phonons}, one can also note an overlap between the acoustic and optical phonon bands. This trend may significantly affect the material's physical properties and behaviors. One crucial implication of this overlap is the potential for phonon-phonon scattering events. When phonons of different types, specifically acoustic and optical, share similar energy ranges, they can interact and scatter off one another. This phenomenon is notable because of its direct impact on the thermal transport properties of the material. 

Our computational protocol also focuses on the thermal stability of PH-Si as we employed AIMD simulations. In these simulations, we employed a $2\times2\times1$ supercell with an NVT ensemble and controlled the temperature using the Nosé-Hoover thermostat. The simulations extend throughout 5 ps, employing a time step of 1.0 fs.

\begin{figure}[!htb]
	\centering
	\includegraphics[width=0.9\linewidth]{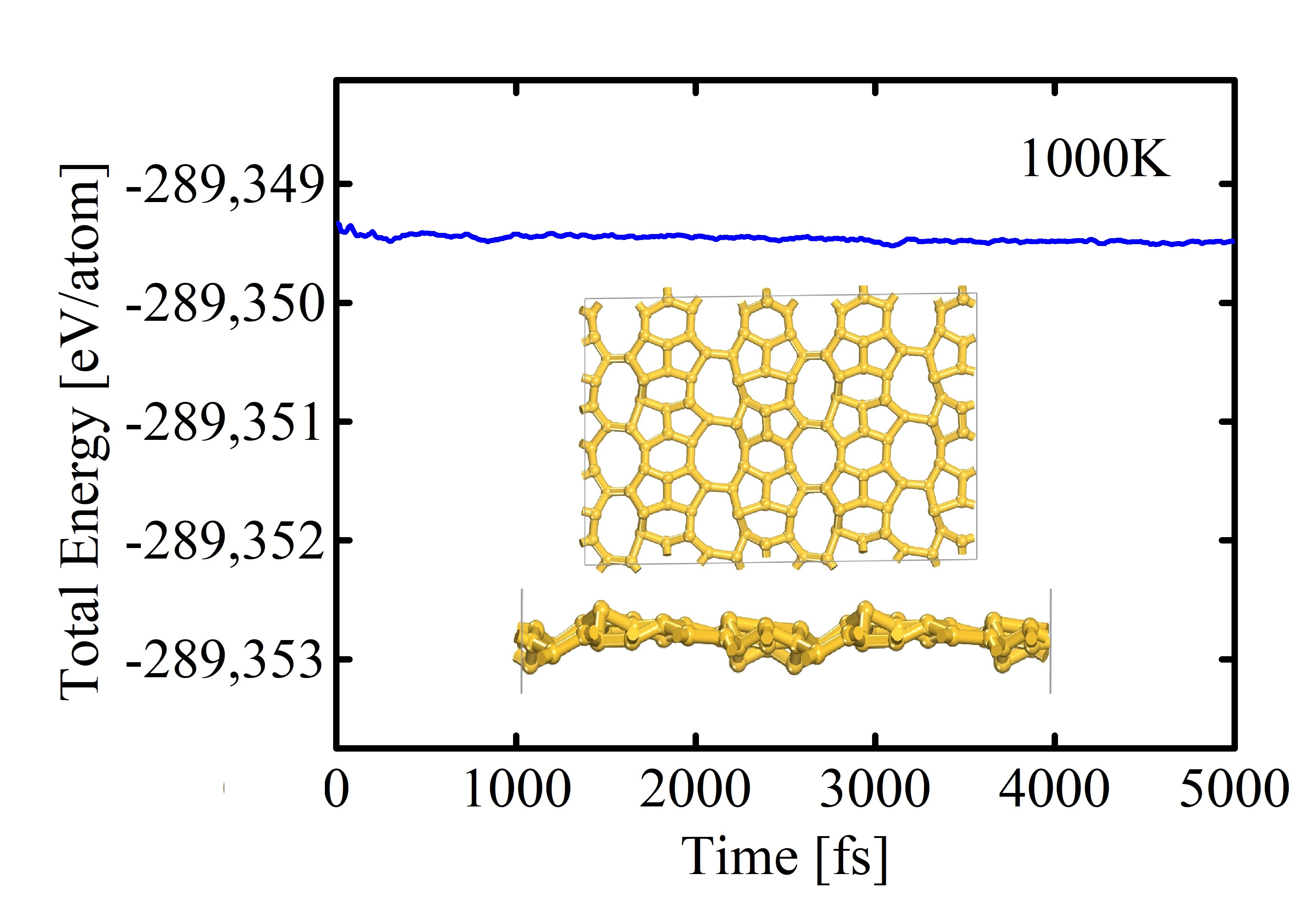}
	\caption{Time evolution of the total energy per atom in the PH-Si lattice at 1000K, using the PBE approach. The insets show the top and side views of the final AIMD snapshot at 5000 fs.}
	\label{fig-aimd}
\end{figure}

To assess the thermal stability of PH-Si, we conducted AIMD simulations. Figure \ref{fig-aimd} presents the temporal evolution of the total energy per atom in PH-Si for a 5000 fs simulation at 1000 K. The snapshots provided in Figure \ref{fig-aimd} capture the final stage of these AIMD simulations.

While comparing the structural features in the snapshots to the optimized lattice shown in Figure \ref{fig:sys}, some degree of deformation in the PH-Si structure becomes evident. However, it is essential to note that there is no indication of bond breakage, and the original configuration remains intact. The deviations observed in the AIMD snapshots primarily relate to alterations in planarity and bond distances among the lattice atoms, which can be attributed to the elevated temperature.

In a thermally stable system, the total energy per atom should exhibit relatively consistent behavior, indicating an even energy distribution within the material. In this context, we observe that the temporal evolution of the total energy displays a nearly flat pattern with minimal fluctuations. This consistency provides further evidence of PH-Si's thermal stability.

\begin{figure}[!htb]
	\centering
	\includegraphics[width=\linewidth]{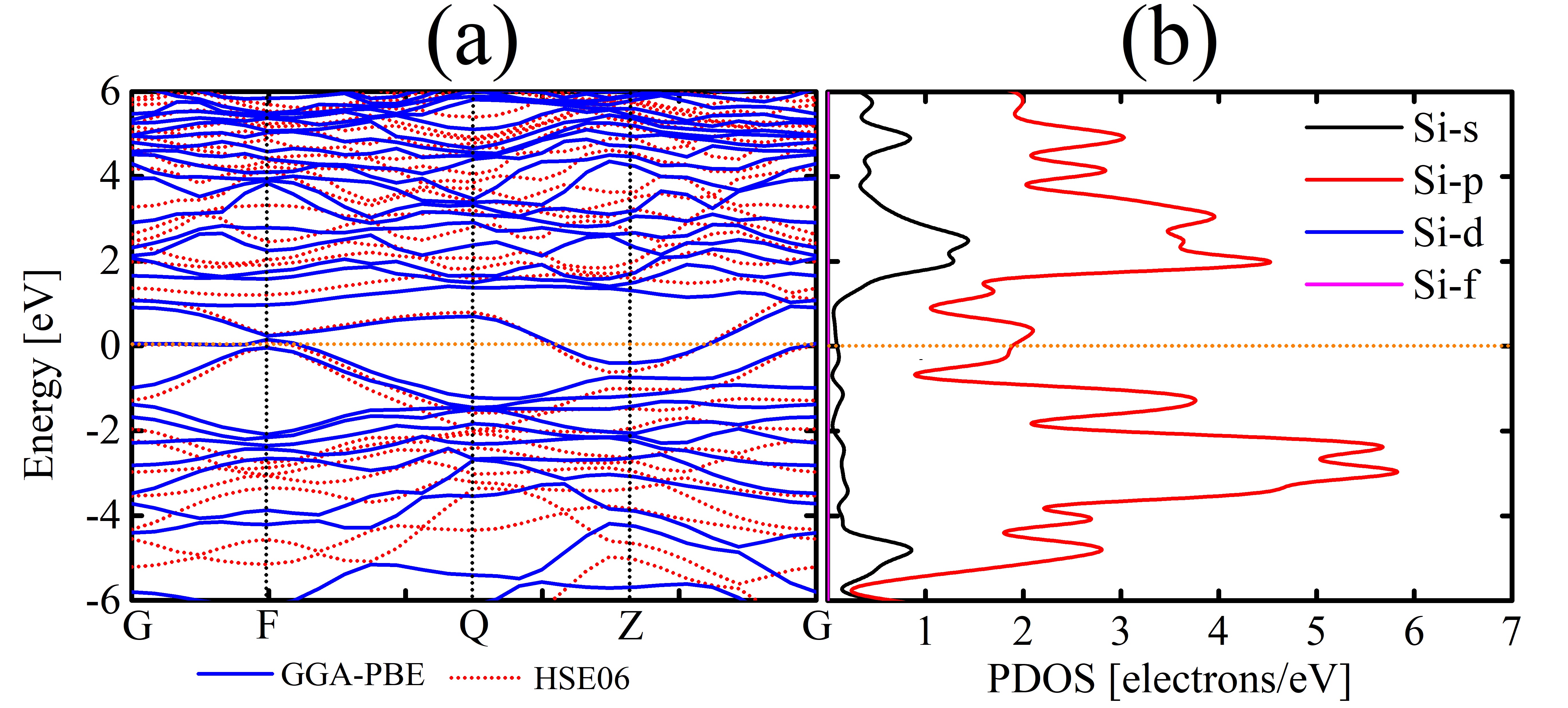}
	\caption{(a) Electronic band structure and (b) partial density of states (PDOS) for PH-Si. The band structure was obtained using the PBE (blue) and HSE06 (red) approaches. PDOS was calculated at the HSE06 level.}
	\label{fig-bands}
\end{figure}

We now discuss the band structure features of PH-Si. Figure \ref{fig-bands} illustrates its (a) electronic band structure and (b) partial density of states (PDOS). The band structure was obtained using the PBE (blue) and HSE06 (red) methods. PDOS was calculated at the HSE06 level. It is worth mentioning that PBE calculations generally underestimate band gaps. Therefore, we also performed HSE06 calculations to determine the electronic and optical properties of PH-Si more accurately. In this way, Figure \ref{fig-bands}(a) clearly shows the pronounced anisotropic conductance within PH-Si. Notably, it exhibits metallic characteristics along the G-F direction while displaying semiconducting behavior along other directions.   

The band structure exhibits linear energy dispersions near the Fermi level in proximity to the F reciprocal point. Additionally, the Dirac point is approximately 0.25 eV above the Fermi level. This positioning suggests that charge carriers within PH-Si can propagate like massless Dirac fermions. This property holds promising potential for applications in nanoelectronics \cite{meindl2001limits}.

One can realize that the PDOS results presented in Figure \ref{fig-bands}(b) unequivocally portray PH-Si as a metal. This feature is evident from the multitude of electronic states at the Fermi. We also observe that the electronic states at the Fermi level primarily emanate from Si-p levels (red line), with a small contribution coming from Si-s levels (black line). This finding suggests that the p-orbitals play a significant role in driving electronic transitions and interactions in PH-Si. These orbitals are often associated with directional bonding phenomena.  

The effective masses of holes ($m_h^*$) and electrons ($m_e^*$) in PH-Si are 0.08$m_0$ and 0.37$m_0$, respectively, where $m_0$ represents the electron mass. Generally, lower effective masses are associated with higher carrier mobility, particularly suitable for microelectronic devices. The $m_h^*$ and $m_e^*$ values for PH-Si are similar to the ones reported for pristine and doped silicene nanosheets and nanostrips \cite{chowdhury2016theoretical,quhe2012tunable,tene2023modeling,fan2018silicene}.

Notably, the $m_e^*$ values for BPN-(Al,Ga)N are lower than those of the typical \textit{wz}-AlN and \textit{wz}-GaN structures (0.22$m_0$ and 0.25$m_0$ \cite{persson2001effective}, respectively). This observation underscores the greater potential of BPN-(Al,Ga)N systems for applications demanding enhanced electronic transport efficiency compared to their wurtzite analogs.

\begin{figure}[!htb]
	\centering
	\includegraphics[width=\linewidth]{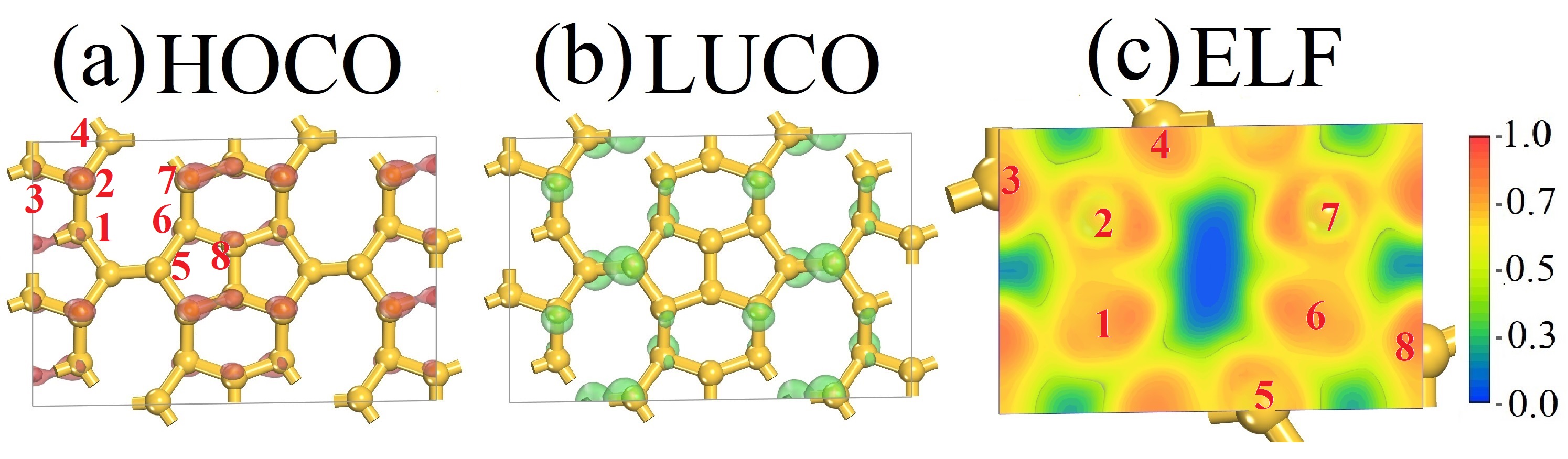}
	\caption{Schematic representation of the (a) highest occupied crystal orbital (HOCO), (b) lowest unoccupied crystal orbital (LUCO), and (c) electron localization function. These properties were calculated using the HSE06 approach.}
	\label{fig-elf}
\end{figure}

For a deeper insight into the underlying chemical interactions within PH-Si, we present the highest occupied crystal orbital (HOCO), lowest unoccupied crystal orbital (LUCO), and electron localization function (ELF) in Figure \ref{fig-elf}. Mainly, ELF helps to understand the bonding status of PH-Si. The localization of HOCO and LUCO in PH-Si are shown in Figures \ref{fig-elf}(a) and \ref{fig-elf}(b), respectively. The HOCO predominantly localizes on Si2, Si3, Si7, and Si8 atoms. In contrast, the LUCO localization is distributed on all the atoms except Si-1 and Si-3. This distribution results in a charge imbalance, where Si1 and Si3 can carry only partial positive charges. 

The ELF depicted in Figure \ref{fig-elf}(c) provides valuable insights into the distribution of electrons within the material. ELF offers a topological perspective on electron interactions, enabling the identification of regions with either electron localization or delocalization. It assigns values between 0 and 1 to each point in space, where values approaching 1 generally indicate strong covalent interactions or the presence of lone pair electrons. In contrast, lower values (around $\sim$0.5) suggest delocalization, ionic bonds, or weak Van der Waals interactions.

Figure \ref{fig-elf} provides a clear depiction of the chemical bond types in PH-Si. Strong $\sigma$ bonds are present between the Si1-Si2 and Si6-Si7 atom pairs, highlighted by intense orange regions with values around 0.7. In contrast, the bonds between other Si atoms are characterized by lighter-yellow areas (with values near 0.5), indicating a considerable degree of electron delocalization. It is well-established that metallic-like conductivity is a hallmark of materials where valence electrons are delocalized and can move freely. Conversely, materials with strong covalent bonds tend to exhibit semiconductor-like conductivity. In the case of PH-Si, the coexistence of localized and delocalized electrons within its bond network underpins the anisotropic conductance observed in its electronic band structure, as discussed earlier. This bond arrangement is pivotal in shaping the electronic properties of PH-Si. 

The optical properties of semiconductor materials are intricately tied to their electronic band structures in the ground state. This relationship stems from the fact that electronic transitions in materials encompass both interband and intraband transitions. These transitions play distinct roles in shaping the optical behavior of semiconductor and metallic materials, where interband transitions are critical for semiconductors, and intraband transitions are dominant in metals.

\begin{figure}[!htb]
	\centering
	\includegraphics[width=\linewidth]{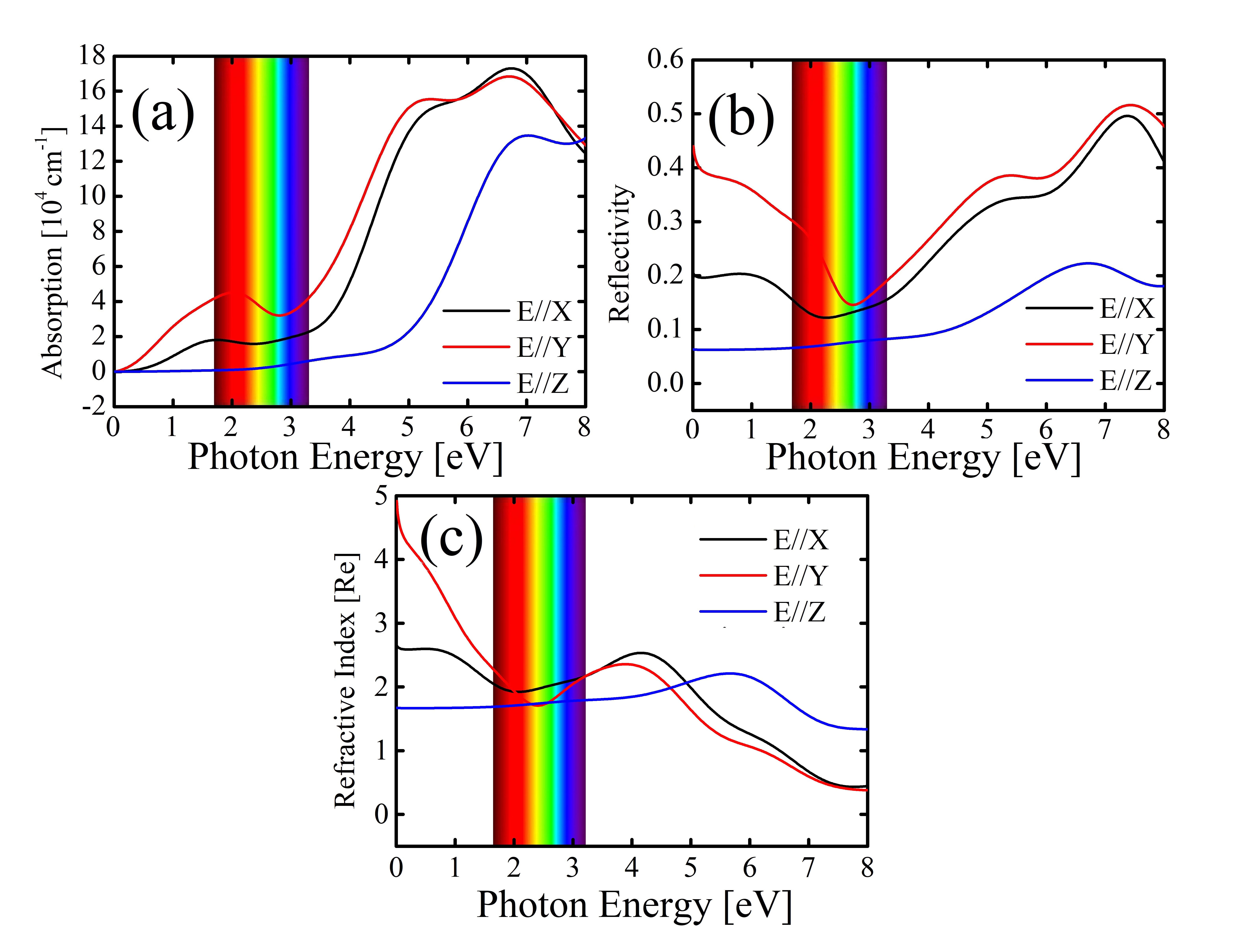}
        \caption{(a) optical absorption, (b) reflectivity, and (c) refractive index calculated at HSE06 level for polarised light beam oriented x (E//X), y (E//Y), and z (E//Z) directions related to the surface oh PH-Si.}
	\label{fig-optical}
\end{figure}

Figure \ref{fig-optical} illustrates the PH-Si's optical characteristics, considering light polarization along the x (E//X), y (E//Y), and z (E//Z) directions. Given its intrinsic electronic anisotropy and structural buckling, we anticipate observing differing properties in the in-plane and out-of-plane polarization directions, at least quantitatively. The optical properties, such as the absorption coefficient, refractive index, and reflectivity, can be derived from the complex dielectric function \cite{lima2023dft}. The real part describes the material's response to incident light, describing its polarization. Conversely, the imaginary part indicates the material's ability to absorb electromagnetic radiation. These quantities can be inferred using the Kramers-Kronig relation \cite{PhysRevB.73.045112}. The optical properties studied here can be computed as described in reference \cite{lima2023dft}.

The optical absorption coefficient quantifies the reduction in light intensity per unit distance as it traverses a medium. Figure \ref{fig-optical}(a) illustrates that PH-Si displays a high absorption coefficient (10$^{4}$ cm$^{-1}$) due to its metallic properties. The first absorption peaks for E//X and E//Y fall within the visible spectrum, similar to observations in silicene \cite{bao2022tuning,chowdhury2016theoretical,matthes2014optical,das2015optical}. However, PH-Si's first peak (approximately 1.9 eV, in the red region of the visible spectrum) exhibits a red-shift of about 0.3 eV compared to silicene, which features its first peak in the green region of the visible spectrum (around 2.2 eV) \cite{bao2022tuning}. 

In s-silicene and PH-Si, intense absorption primarily occurs within the ultraviolet (UV) region \cite{bao2022tuning,chowdhury2016theoretical,matthes2014optical,das2015optical}. PH-Si shows absorption peaks of about 17.0$\times$10$^{4}$ cm$^{-1}$ for E//X and E//Y and 13.0$\times$10$^{4}$ cm$^{-1}$ for E//Z for a photon energy of 6.8 eV. For silicene, the maximum absorption coefficient reaches 13.5$\times$10$^{4}$ cm$^{-1}$ at a photon energy of 4.07 eV under in-plane light polarization. In the case of out-of-plane light polarization, it peaks at 16.35$\times$10$^{4}$ cm$^{-1}$ with a photon energy of 9.11 eV, both falling within the ultraviolet (UV) range \cite{das2015optical}. These results suggest that PH-Si, similarly to silicene \cite{das2015optical}, has potential applications as Vis-UV detectors and absorbers. 

In Figure \ref{fig-optical}(b), PH-Si exhibits relatively low to moderate reflectivity (less than 0.30 and 0.52) across the visible to ultraviolet (Vis-UV) regions. It is important to note that the reflection function signifies the ratio of photon energy reflected off the surface to that incident upon the surface. The highest reflection coefficient observed for PH-Si is 0.51 at a photon energy of 7.3 eV. Moreover, incident light in PH-Si is efficiently transmitted in the frequency range that spans the visible region. The material displays minimum reflection coefficients of about 0.13 and 0.07 for in-plane and out-of-plane light polarizations, respectively. Notably, PH-Si exhibits some reflectivity peaks in the infrared (IR) region for in-plane light polarization. Given its relatively low light absorption within this range, it suggests that PH-Si can serve as an IR protector. It is worth mentioning that silicene exhibits similar optical characteristics \cite{bao2022tuning}.

Birefringence in materials is observed when the velocity of light differs in distinct polarization directions. This phenomenon is quantified as the disparity between the extraordinary and ordinary refractive indices, i.e., the index of refraction for electric fields oriented parallel and perpendicular to the materials plane, respectively \cite{john2017optical}. In PH-Si, the refractive indices along parallel and perpendicular polarization directions to its basal plane display anisotropy, as illustrated in Figure \ref{fig-optical}(c). This trend suggests that PH-Si is a birefringent material, similar to silicene \cite{john2017optical}.

Finally, we analyze the elastic properties of PH-Si. An evaluation of the material's anisotropy in mechanical properties is carried out by determining Poisson's ratio ($\nu(\theta)$) and Young's modulus ($Y(\theta)$) under pressure within the xy plane. These properties are derived from the following equations \cite{doi:10.1021/acsami.9b10472,doi:10.1021/acs.jpclett.8b00616}:

\begin{equation}
    \displaystyle Y(\theta) = \frac{{C_{11}C_{22} - C_{12}^2}}{{C_{11}\alpha^4 + C_{22}\beta^4 + \left(\frac{{C_{11}C_{22} - C_{12}^2}}{{C_{44}}} - 2.0C_{12}\right)\alpha^2\beta^2}}
    \label{young}
\end{equation}

\noindent and 

\begin{equation}
    \displaystyle \nu(\theta)= \frac{{(C_{11} + C_{22} - \frac{{C_{11}C_{22} - C_{12}^2}}{{C_{44}}})\alpha^2\beta^2 - C_{12}(\alpha^4 + \beta^4)}}{{C_{11}\alpha^4 + C_{22}\beta^4 + \left(\frac{{C_{11}C_{22} - C_{12}^2}}{{C_{44}}} - 2.0C_{12}\right)\alpha^2\beta^2}}.
    \label{poisson}
\end{equation}

\begin{figure}[htb!]
	\centering
	\includegraphics[width=\linewidth]{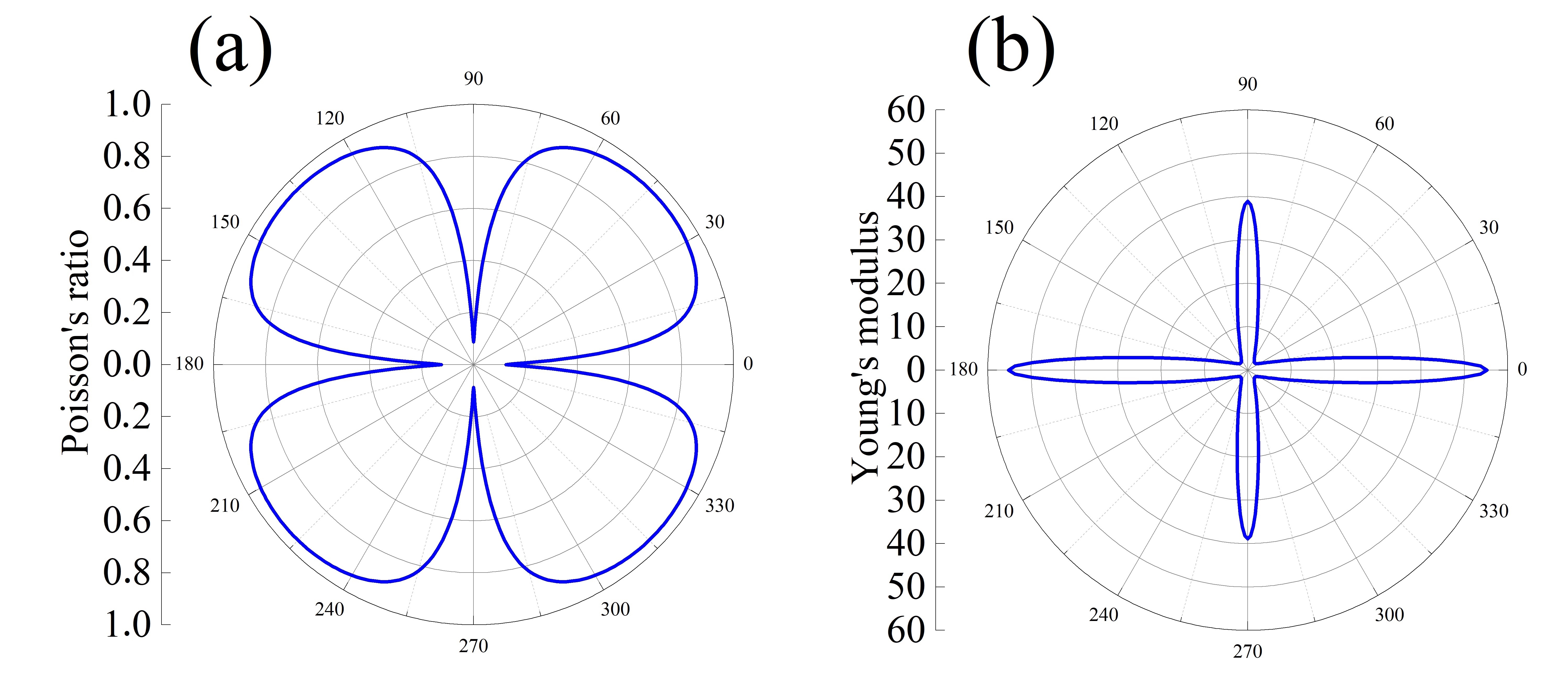}
	\caption{2D representation of (a) Poisson’s ratio and (b) Young's modulus in the basal plane for PH-Si.}
	\label{fig:elastic}
\end{figure}

\noindent In the equations above, $\alpha=\cos(\theta)$ and $\beta=\sin(\theta)$. To provide a complete understanding of the mechanical properties of PH-Si, we have summarized the elastic constants and parameters in Table \ref{tab:elastic}. Furthermore, we visualized the distribution of Poisson's ratio (Figure \ref{fig:elastic}(a)) and Young's modulus (Figure \ref{fig:elastic}(b)) values across its basal plane. These mechanical properties are anisotropic in PH-Si. Importantly, the elastic constants in Table \ref{tab:elastic} agree with the Born-Huang criteria for orthorhombic crystals ($C_{11}C_{22} - C_{12}^2>0$ and $C_{44}>0$) \cite{PhysRevB.90.224104,doi:10.1021/acs.jpcc.9b09593}, suggesting the good mechanical stability of PH-Si.

\begin{table}[!htb]
\centering
\caption{Elastic constants C$_{ij}$ (GPa) and maximum values for Young's modulus (GPa) ($Y_{MAX}$) and maximum ($\nu_{MAX}$) and ($\nu_{MIN}$) Poisson's ratios.}
\label{tab:elastic}
\begin{tabular}{| l |c|c|c|c|c|c|c|c|}
\hline
 Structure & C$_{11}$ & C$_{12}$ &C$_{22}$ &C$_{44}$ & $Y_{MAX}$  & $\nu_{MAX}$ & $\nu_{MIN}$ \\
 \hline
PH-Si    & $ 39.33$       & $4.89$     & $55.81$     & $0.59$  & $ 38.05$ & $0.81$ & $0.14$        \\
 \hline
 \end{tabular}
\end{table}

For Young's modulus, we obtained a maximum value ($Y_{MAX}$) of 38.05 GPa, almost half of the value reported for (61.7 GPa \cite{mortazavi2017first}). This discrepancy can be attributed to the inherent porosity shown by PH-Si, which arises from the presence of eight-atom rings and the rigidity of the bonds in fused pentagonal rings. These factors collectively contribute to the reduced strain resilience of PH-Si compared to its hexagonal counterpart (silicene).

When subjected to a compressive (tensile) strain in one direction, materials typically exhibit expansion (contraction) in the perpendicular direction, resulting in positive Poisson's ratios, which is the case with most ordinary materials. In Figure \ref{fig:elastic}, one can observe that PH-Si shows positive Poisson's ratios, with a maximum value ($\nu_{MAX}$) of approximately 0.81. For this value, Young's modulus values remain below 10 GPa, indicating the incompressibility of these materials under biaxial strains.

Common materials typically exhibit Poisson's ratios that fall within the range of 0.2 to 0.5 \cite{greaves2011poisson}. A Poisson's ratio of 0.5 characterizes incompressible materials. i.e., their lateral dimensions are not changed when subjected to axial strain. The minimal Poisson's ratio ($\nu_{MIN}$) in PH-Si under uniaxial strains along the x and y directions are 0.21 and 0.14, respectively. These values are notably lower than those observed in silicene, which are approximately 0.33 and 0.29 \cite{mortazavi2017first}, respectively. The reduced Poisson ratios in PH-Si are also a consequence of its lattice arrangement, which exhibits a higher degree of porosity when compared to its honeycomb-based lattice counterpart (silicene). This increased porosity allows PH-Si to undergo more deformation under tension, resulting in lower Poisson ratios.

\section{Conclusion}

In this study, we have introduced a novel 2D silicon allotrope, pentahexoctite silicon (PH-Si), inspired by the unique structural features of pentahexoctite carbon. Our comprehensive investigation has shed light on the fundamental properties of this material.

The bond lengths in PH-Si suggest that this material may exhibit stability on par with silicene. The Si-Si bond lengths in PH-Si are also similar to those found in pristine pentagonal-ring-based silicon materials. In addition, like other 2D silicon materials, PH-Si also exhibits buckling characteristics. Our analysis has also unveiled that its formation energy is about -3.62 eV.

PH-Si exhibits intriguing electronic characteristics, featuring anisotropic conductance with metallic behavior along specific directions, making it a potential candidate for nanoelectronic applications. Moreover, PH-Si demonstrates strong optical absorption in the visible and ultraviolet regions, suggesting its suitability as a visible-ultraviolet detector and absorber. The mechanical properties highlighted the anisotropic behavior of PH-Si, including its Poisson's ratio (0.14-0.81) and Young's modulus (10.0-38.5 GPa).  

\section*{Acknowledgement}

This work was financed by the Coordenação de Aperfeiçoamento de Pessoal de Nível Superior (CAPES), Conselho Nacional de Desenvolvimento Cientifico e Tecnológico (CNPq), and Fundação de Apoio à Pesquisa do Distrito Federal (FAP-DF). L.A.R.J. acknowledges the financial support from FAP-DF $00193.00001808/2022-71$ grant and FAPDF-PRONEM grant $00193.00001247/2021-20$, and CNPq grant $350176/2022-1$. L.A.R.J. acknowledges N\'ucleo de Computaç\~ao de Alto Desempenho (N.A.C.A.D.) and for providing the computational facilities. This work used resources of the Centro Nacional de Processamento de Alto Desempenho em São Paulo (CENAPAD-SP). L.A.R.J. and K.A.L.L. also acknowledge CAPES for partially financing this study - Finance Code 88887.691997/2022-00.

\bibliographystyle{unsrt}
\bibliography{bibliography.bib}

\end{document}